\documentclass[12pt,a4paper]{article}

\usepackage[a4paper]{geometry}
\newcommand{\dd}{\partial}

\newcommand{\ks}{k_{*}}

\newcommand{\beq}{\begin{equation}}
\newcommand{\eeq}{\end{equation}}
\usepackage{amsmath,graphicx,amssymb}

\title{What is isotropic turbulence and why is it important?} 
\author{David McComb\\
SUPA School of Physics and Astronomy,\\
Peter Guthrie Tait Road, \\
University of Edinburgh,\\
EDINBURGH EH9 3JZ.\\
Email: wdm@ph.ed.ac.uk \\
Website and blog: blogs.ed.ac.uk/physics-of-turbulence/}
\date{}
\begin{document}
\maketitle
\thispagestyle{empty}
\begin{abstract}
This article begins with an overview, then gives the precise definition of isotropic turbulence, and follows that with the basic conservation equations, in both real space and wavenumber space. These provide the foundations of all theoretical approaches, both fundamental and phenomenological. After that, my intention is to try to highlight the main unresolved issues and give some indication of what progress there has been over decades (in all cases), and what still needs to be done. I should emphasise that I am not trying to provide either a conventional review or even a pedagogical treatment. Instead I am giving concise summaries, supplemented (where I can) by my own observations, which make substantial points that I believe are original, and which have not been made in the literature. To take just one example, it is known by some people that Kolmogorov's 1962 theory is not correctly described as a `refinement' of his 1941 theory. This was pointed out by Kraichnan in 1974. However, what does not appear to have been recognized is that the 1962 theory is physically invalid, and also that a plausible implementation of it destroys the Kolmogorov (1941) scaling of  energy spectra which has been widely observed over many years. This is discussed in Section 4 below. Lastly, I have tried to give an informal treatment in order to make everything easily accessible, to reach the widest possible audience. In particular, the section on renormalization methods is written without giving the  equations of the various theories, merely stating in words what has been done, what are the different methods and also what still needs to be done.
\end{abstract}
\newpage

\tableofcontents
\newpage

\section{Introduction}
 
Near the end of the nineteenth century in Manchester, Osborne Reynolds studied turbulent flow through pipes; and, in the process, laid the foundations of research in turbulence for the next century and more. His introduction of a dimensionless criterion for the existence of turbulence, the Reynolds number, is well known; as is his introduction of the concept of the Reynolds stress due to the correlation of fluctuating velocities. Furthermore, he also discovered the fundamental problem of the statistical theory of turbulence, which is \emph{the closure problem}. He found that, in using the Navier-Stokes equations (NSE) to derive an equation for the mean velocity,  one introduces the covariance of two fluctuating velocities. This leads to an  open hierarchy of moment equations, with the equation for the $n$-order moment containing the unknown $(n+1)$-order moment, and so on.

It is illustrative to consider what this means for one of the simplest possible practical situations, which is (in the spirit of Reynolds) flow under applied pressure through a straight pipe of circular cross-section. If first we consider laminar Poiseuille flow along a pipe, then we may obtain an exact solution for the velocity distribution in terms of the applied pressure. This is because the nonlinear term vanishes exactly; and, in the continuum limit, the \emph{microscopic} statistical closure problem has been solved with great accuracy, by the introduction of the coefficient of viscosity of the macroscopic fluid. If we then consider the analogous situation of turbulent Poiseuille flow (as based on the mean velocity) along a pipe, the nonlinear term vanishes for the mean flow, but we still have the occurrence of the covariance of fluctuating velocities through the Reynolds stress. Thus, until we can solve the closure problem, we cannot solve the turbulent Poiseuille flow problem in the way that we can do for the laminar case.

Historically, this problem has been treated by the \emph{ad hoc} introduction of an effective viscosity due to the turbulence, with analogies being drawn between the randomizing effect of the turbulent eddies and the molecular motions of the microscopic case. However, such analogies are rather crude, as the turbulent motions  involve a continuous range of scales whereas taking the continuum limit in the molecular case involves a huge scale separation. We will return later to the many more fundamental approaches to this problem, when we restrict our attention to isotropic turbulence. However, for the moment we would like to generalise this observation to all comparisons between turbulence research and mainstream physics. It is important to bear in mind that the number of degrees of freedom in a typical turbulent flow is of the order thousands, whereas the number of degrees of freedom $N$ in, for instance, the molecular description of a cubic centimetre of air is of the order of one billion billion. This means that the statistical error, which is of order $1/\surd{N}$, is too small to be measured in macroscopic physics.

\subsection{Is turbulence research still in its infancy?}

Around about the end of the twentieth century, two major review articles attempted to give an overview of the work done during that century. In 1999,  Sreenivasan reviewed the study of fluid turbulence \cite{Sreenivasan99}. His emphasis was on the difficulty of taking a unified view of the subject, particularly because of the diversity of its many applications; and also its possible relationship to various other activities, such as chaos theory, fractals or the theory of critical phenomena. He made the point that the quantum theory of a single-electron atom cannot be straightforwardly  applied to a multielectron atom: here complexity comes into the picture, and one is forced to consider an approximate solution, which is obtained by an approximate procedure  (such as mean-field theory). This is generally true in physics where complexity or strong coupling (or both) is involved, we can only hope for approximate solutions in general. Hence, in making comparisons, turbulence should be compared to the more complicated problems in physics: in fact, it is an example of a many-body  problem.

Sreenivasan also raised a number of questions about various sub-topics of turbulence research. From our present point of view, the need for an understanding of the effects of finite Reynolds number is probably the most important of these.

In 2001, Lumley and Yaglom published an article with the title \emph{A century of turbulence} \cite{Lumley01}. After pointing out that neither author had personally taken part in research on turbulence for the entire century (and also acknowledging that their title might be misinterpreted by many as relating to political matters and international affairs!) they made a comprehensive, almost encyclopaedic, examination of the history of the study of turbulence. They stated their conclusions in their Abstract, as follows:
\begin{quote}
`This field does not appear to have a pyramidal structure, like the best of physics. We have very few great hypotheses. Most of our experiments are exploratory experiments. What does this mean?'
\end{quote}
They went on to answer their own question:
\begin{quote}
`We believe it means that, even after 100 years, turbulence studies are still in their infancy.'
\end{quote}
I am not quite sure what is meant by the phrase `pyramidal structure' in the context of physics. Presumably they were thinking in terms of the taxonomy of living things, where the taxonomy of animals (for instance) can be represented by a pyramidal structure. But overall the general sense is clear; and really quite persuasive. They find a lack of coherence and unifying principles about the whole subject, and this is also the broad conclusion of Sreenivasan. Indeed, even after a further two decades which have been marked by an explosive growth in research, this depressing view is still to a considerable view justified.

Sreenivasan (\emph{ibid}) ends up (in effect) by asking the question:
\begin{quote}
`How may one recognize  that the turbulence problem has been solved?'
\end{quote}
We shall return to that question, but here we may simply note that it is first necessary to have a consensus on what actually \emph{is} the turbulence problem. And also, perhaps, an acceptance that `solved' means `solved \emph{approximately} to some acceptable level of accuracy'.

\subsection{A pragmatic taxonomy for turbulence}

So far as I am aware, there is no formal hierarchical classification of turbulent flows. Nevertheless, when I began my postgraduate studies in 1966 there was an informal understanding in the field that flows could be classified in terms of their degee of symmetry. From a practical point of view, this was equivalent to classifying them in terms of how many independent variables were needed to specify the flow. Evidently, for computation or measurement, the fewer the better. Thus, at the top of our notional pyramid,would be isotropic turbulence, with homogeneous turbulence immediately below it, followed by a vast and messy array of more realistic or practical problems!

Isotropic turbulence may be seen as turbulence reduced to its essentials. It is, in effect, turbulence as physics. For this reason, irrespective of practical considerations, the fundamental problem of turbulence is the statistical closure problem posed by isotropic turbulence. By restricting our attention to isotropic turbulence, we exclude many of the considerations that made the authors of the reviews \cite{Sreenivasan99},\cite{Lumley01} so downbeat. That is why it is time to concentrate a critical examination on isotropic turbulence as we are doing in this Special Issue. 

In the rest of this Editorial, we shall first set the scene, and then concentrate on more technical matters. All the notation used will be found in the book \cite{McComb14a}.

\section{Definition of isotropic turbulence}

First we have to say what we mean by homogeneous turbulence. This is defined very clearly , on page 1 of the book by Batchelor \cite{Batchelor71} as:
\begin{quote}
`Our concern is with homogeneous turbulence, which is a random motion whose \emph{average} properties are independent of position in the fluid.'
\end{quote}
Note that the emphasis in this, and the next, definition is mine.

Batchelor then pointed out that a greater simplification could be achieved by additionally restricting the turbulence to be \emph{statistically isotropic} as well. He also commented that:
\begin{quote}
The possibility of this further assumption of isotropy exists only when the turbulence is already homogeneous, for certain directions would be preferred by a lack of homogeneity.
\end{quote}
Thus homogeneity is a necessary, but not sufficient, condition for isotropy.

This means that when when we refer to `homogenous, isotropic turbulence', the first word is redundant. Of course we shall all go on referring to it that way and using the three-letter acronym \emph{HIT}. Nevertheless, it is essential to be aware of the distinction for a proper understanding of the subject.

It is important to remind ourselves of these definitions. I realise that for many people it will be trivially obvious that it is impossible to test the random velocity field itself for properties such as homogeneity or isotropy. However, remarks crop up in the literature to the effect that `intermittency at the small scales' (\emph{sic}) destroys homogeneity. This is impossible, because intermittency is a single-realization phenomenon and, as we have just seen, homogeneity is a property of average quantities, where the sum over all the realizations will have averaged out the intermittency. One should also note that numerical simulations have shown that internal intermittency is a property associated with \emph{all} scales of turbulence.

\subsection{The implications of isotropy for the statistical formulation}

Let us denote the random velocity field by $u_{\alpha}(\mathbf{x},t)$, where the index $\alpha$ is the usual Cartesian tensor index and takes values $\alpha = 1,\,2\, \mbox{or} \,3$. The use of Greek letters for subscript indices frees up letters like $k,j,l,n,m \dots$ for wavenumbers in theoretical work. This usage actually dates back to Kolmogorov in 1941.

It is usual in shear flows to be interested in the mean velocity, but the first consequence of imposing isotropy is that the mean velocity must vanish. That is, we must have:
\beq
\langle u_{\alpha}(\mathbf{x},t)\rangle =0,
\eeq
where $\langle \dots \rangle$ denotes the ensemble average. This applies to the average of any vector quantity. As a result, we concentrate on the mean of the square of the velocities, along with related quantities such as correlations, structure functions and energy spectra. Accordingly, we introduce the pair-correlation of velocities in its most general form $C_{\alpha\beta}$ as:
\beq
C_{\alpha\beta}(\mathbf{x},\mathbf{x};t,t')=\langle u_{\alpha}(\mathbf{x},t)u_{\beta}(\mathbf{x'},t')\rangle.
\label{gen_corr}
\eeq
For a general approach, we introduce the centroid and relative coordinates, thus:
\beq
\mathbf{R} = (\mathbf{x} + \mathbf{x'})/2; \qquad \mbox{and} \qquad\mathbf{r}=(\mathbf{x'}-\mathbf{x});
\eeq
and similarly for the time:
\beq
T = (t + t')/2; \qquad \mbox{and} \qquad \tau=(t'-t).
\eeq
Hence we may write the pair correlation as:
\beq
C_{\alpha\beta}(\mathbf{r},\mathbf{R};\tau,T)=\langle u_{\alpha}(\mathbf{x},t)u_{\beta}(\mathbf{x'},t')\rangle.
\label{mod_gen_corr}
\eeq

This is the most general form and applies to any flow configuration. Here we want to specialise it to the isotropic case, and of course we begin with homogeneity, where there is no dependence on the centroid coordinate, thus:
\beq
C_{\alpha\beta}(\mathbf{r},\mathbf{R};\tau,T) \rightarrow C_{\alpha\beta}(\mathbf{r};\tau,T);
\eeq
and similarly for stationarity (which is homogeneity in time), we may make the further reduction:
\beq
C_{\alpha\beta}(\mathbf{r},\mathbf{R};\tau,T) \rightarrow C_{\alpha\beta}(\mathbf{r};\tau).
\eeq
For isotropic turbulence we may anticipate that the problem can be reduced further by introducing a scalar function $C(r,\tau)$ by
\beq
tr\, C_{\alpha\beta}(\mathbf{r};\tau) = 2C(r,\tau).
\eeq
Note that in order to pursue a theoretical approach or carry out a numerical simulation we normally  Fourier transform these results with respect to $r$ to obtain the forms in wavenumber space as functions of $k$. Less frequently the same operation may be performed for the time $t$ and angular frequency $\omega$. Full dicussions of these matters may be found in the book \cite{McComb14a}, but for later convenience we give two results here. For isotropic turbulence, the Fourier-transformed two-point, two-time covariance becomes:
\beq
C(k;t,t');
\label{spec_density}
\eeq
and this is related to the energy spectrum when $t=t'$, by:
\beq
E(k,t) = 4\pi k^2 C(k;t,t).
\label{en_spect}
\eeq

\section{The energy balance equations in $x$-space and $k$-space} 

We take the fluid motion to be governed by the Navier-Stokes equations (NSE, for short), which express conservation of momentum, and we note that for isotropic turbulence that it averages term-by-term to zero\footnote{Non-zero averages of vector quantities would be symmetry-breaking in an isotropic field.}. In order to formulate a statistical treatment therefore we have to work with second-order statistical quantities such as the moments, structure functions and spectra of the velocity field.

\subsection{The Karman-Howarth equation}

We begin with the Karman-Howarth equation (KHE) in $x$-space, which relates the second- and third-order moments. This expresses conservation of energy in real space  for any specific value of the relative coordinate $r=x'-x$. In physics the term for it is a \emph{local energy balance equation}. Originally Von Karman and Howarth derived it in terms of the second- and third-order correlation functions \cite{VonKarman38}. In fact we state the form of the KHE equation when it is written in terms of the second-order and third-order structure functions, $S_2$ and $S_3$,  thus:
\beq
0=-\frac{2}{3}\frac{\dd E}{\dd t} + \frac{1}{2}\frac{\dd S_2}{\dd t} + \frac{1}{6r^4}\frac{\dd}{\dd r}(r^4 S_3) - \frac{\nu}{r^4}\frac{\dd}{\dd r}\left(r^4\frac{\dd S_2}{\dd r}\right),
\label{khe}
\eeq
as this form will be helpful when we consider the Kolmogorov theories later on.  Also, I have moved the term involving the total energy (per unit mass) $E$ to the right of the equals sign, for a reason which will become obvious.

\subsection{The Lin equation}

Turning now to a spectral description, we may introduce the Lin equation, which is just the Fourier transform of the KHE. This also can be described as a local energy balance, but this time in wavenumber space. Unfortunately, the real problem with this usage is that it is only valid for the simplest form of the Lin equation, and that in itself can be misleading. We will return to this point presently.

Let us now consider the Lin equation in terms of the energy spectrum $E(k,t)$ and the transfer spectrum $T(k,t)$. We may write it in its well-known form:
\beq
\frac{d E(k,t)}{dt} = T(k,t) - 2\nu k^2E(k,t) \equiv T(k,t) - D(k,t),
\label{lin}
\eeq
where $D(k,t)$ is the energy dissipation spectrum, and comparison of the two forms of Lin equation gives us: $D(k,t) = 2\nu k^{2}E(k,t)$.
Here, as with the KHE, we assume that there are no forces acting. However, unlike with the KHE, this is not the whole story. We may also express the transfer spectrum in terms of its spectral density $S(k,j;t)$ thus:
\beq
T(k,t) = \int_0^\infty\, dj \,S(k,j;t),\quad \mbox{where} \quad  S(k,j;t) = -S(j,k;t);
\label{Tdef}
\eeq
and $S(k,j;t)$ contains the triple moment in wavenumber space: see \cite{McComb14a}. Note that the antisymmetry of $S(k,j;t)$ under the interchange of $k$ and $j$ guarantees that conservation of energy is maintained in the form $\int_0^\infty\, dk T(k) = 0$.

When we substitute for $T(k)$ in terms of $S(k,j;t)$, we obtain the second form of the Lin equation. this is actually more comparable with the KHE, as given above, because the transfer spectrum density contains the Fourier transform of the third-order moment, which of course is always shown explicitly in the KHE\footnote{If only in the form of the third-order structure function or  correlation function.}.

Now compare the two equations. The KHE holds for any value of the independent variable. If we take some particular value of the independent variable, then each term can be evaluated as a number corresponding to that value of $r$, and the above equation becomes a set of four numbers adding up to zero. If we consider another value of $r$, then we have a different four numbers but they must still add up to zero. In short, KHE is local in the independent variable.

In contrast, the Lin equation, if we write it in its full form, tells us that all the Fourier modes are coupled to each other. It is, in the language of physics, an example of the \emph{many-body problem}. It is in fact highly non-local, as in principle it couples every mode to every other mode.

A corollary of this is that the KHE does not predict an energy cascade, but the Lin equation does. This can be deduced from the nonlinear term which couples all modes together plus the presence of the viscous term which is symmetry-breaking. If the viscous term were set equal to zero, then the coupled but inviscid equation would yield equipartition states.

We may consider the transfer of energy from wavenumbers less than $\kappa$ to wavenumbers greater than $\kappa$. To do this, we integrate the terms of the Lin equation from $k=0$ to an arbitarily chosen $k=\kappa$, with the result:
\begin{eqnarray}
\frac{d}{dt}\int^{\kappa}_{0}dk E(k,t) & = &\int_{0}^{\kappa}dk\int^{\infty}_{0}dj S(k,j;t)-\int^{\kappa}_{0}dk D(k,t)  \nonumber \\
& = & \int_{0}^{\kappa}dk\int^{\infty}_{\kappa}dj S(k,j;t)-\int^{\kappa}_{0}dk D(k,t),
\label{linint}
\end{eqnarray}
where the second form of the right hand side relies on the fact that the double integral over $ 0\leq k,j \leq \kappa $ vanishes due to the antisymmetry of $S(k,j;t)$. Evidently this equation tells us that the loss of energy from modes with $k\leq \kappa$ is due to transfer to modes with $k\geq \kappa$, as well as the direct loss to dissipation. We will consider this further in the next section.

As regards the origins of the Lin equation, Eyink and Sreenivasan \cite{Eyink06} have traced its derivation to a letter from Lin to Onsager in 1945, and it has become quite usual to refer to it as the  Lin equation (e.g. see Sagaut and Cambon \cite{Sagaut08}). However, it is interesting to note that an equation apparently identical to (\ref{lin}) appears as equation (9) in 1941 in a paper by Obukhov \cite{Obukhov41}, where it is described as `a balance equation'. In fact, on closer inspection, it turns out to be the integrated form of the Lin equation, which we have as (\ref{linint}); but rather confusingly employs exactly the same notation as the Lin equation.

\subsection{Inertial transfer and Onsager's observation}

In order to consider the inertial transfer further, we first introduce the symbol $E_{\kappa}(t)$ for the amount of energy contained in modes $0\leq k \leq \kappa$, thus:
\beq
\frac{d}{dt}E_{\kappa}(t) = \frac{d}{dt}\int^{\kappa}_{0}dk E(k,t).
\eeq
We then introduce the symbol $\Pi(\kappa,t)$, defined by:
\beq
\Pi(\kappa,t) = \int^{\infty}_{\kappa}dk T(k,t) = -\int^{\kappa}_{0}dk T(k,t).
\label{Pidef}
\eeq
which represents the flux through mode $k=\kappa$; and, in terms of the transfer spectral density,
\beq
\Pi(\kappa,t) = \int^{\infty}_{\kappa}dk\int^{\kappa}_{0}dj\,S(k,j;t) = -\int^{\kappa}_{0}dk\int^{\infty}_{\kappa}dj\, S(k,j;t).
\eeq
Accordingly, we may write the partially integrated Lin equation (\ref{linint}) as:
\beq
\frac{d}{dt}E_{\kappa}(t) = -\Pi(\kappa,t) - \int^{\kappa}_{0}dk\,D(k,t).
\eeq
Thus, the energy contained in modes $0\leq k \leq \kappa$ is lost partly to direct viscous dissipation and partly due to transfer higher-wavenumber modes.This is the only concept of localness which is required for the Richardson-Kolmogorov picture in wavenumber space (K41).

This analysis holds for any wavenumber $k=\kappa$, but the most important case occurs at  $\kappa =\ks$, which is the wavenumber where $T(k,t)$ has its  single zero-crossing. As is well known, the zeros of the transfer spectrum are given by:
\beq
T(0,t) =0;\qquad T(\ks,t)=0; \quad \mbox{and} \quad \lim_{k \to \infty} T(k,t)=0:  
\eeq
while by simple calculus, the behaviour of the energy flux is given by: 
\beq
\Pi(0,t) =0;\qquad \Pi(\ks,t)=\Pi_{max}; \quad \mbox{and} \quad \lim_{k \to \infty} \Pi(k,t)=0.  
\eeq
If follows from conservation of energy that the maximum value that $\Pi_{max}$ can take is the rate of viscous dissipation, thus we have the general result:
\beq
\Pi_{max} \leq \varepsilon;
\eeq
where the equality applies if the \emph{local} viscous dissipation can be neglected at $k=\ks$.

For many years, arising from Onsager's observation in 1945 \cite{Onsager45}, the condition $\Pi_{max} = \varepsilon$ for a range of wavenumbers $k_{bot}\leq k \leq k_{top}$ has been seen as a criterion for the existence of an inertial range, and hence for the Kolmogorov $-5/3$ spectrum holding over that range. However, a numerical investigation by Shanmugasundaram, which was based on the Local Energy Transfer (LET) theory \cite{Shanmugasundaram92}, does not seem to support this conventional picture, which amounts to scale-invariance of the energy flux. For Taylor-Reynolds numbers $4.7\leq R_{\lambda} \leq 254$, the energy flux $\Pi$ was found to take the peaked form predicted from the general behavioural arguments given above, with the peak occurring at $k=\ks$; while, as the Reynolds number increases, the energy spectrum tends to the $-5/3$ form. For the particular case of $R_{\lambda} = 254$, the energy spectrum in Fig. 3 shows at least a decade of $k^{-5/3}$, as one would expect; whereas, from Fig. 6 we see that the energy flux, corresponding to the energy balance in the upper panel of the figure, takes the form of a peak, with $\Pi_{max}/\varepsilon$ lying between 0.70 and 0.80, rather than the value of unity that Onsager suggested. This would seem to be a good example of what prompted Kraichnan's comment \cite{Kraichnan74}:
\begin{quote}
`Kolmogorov's 1941 theory has achieved an embarrassment of success.'
\end{quote}
In other words, despite the underlying conditions not apparently being satisfied, the $-5/3$ spectrum is still observed.

\subsection{The scale-invariance paradox}

This turns out to be one aspect of an apparent contradiction or \emph{paradox}. It may be resolved as follows. Let us return to equation (\ref{Pidef}) which defines the flux, and concentrate on the second of the two expressions on the right hand side, thus:
\beq
\Pi(\kappa,t) = -\int^{\kappa}_{0}dk T(k,t).
\eeq 
Then we substitute from (\ref{Tdef}) in terms of the transfer spectral density:
\begin{eqnarray}
\Pi(\kappa,t)& = & -\int^{\kappa}_{0}dk \int_0^{\infty}dj S(k,j;t) =-\int^{\kappa}_{0}dk \int_0^{\kappa}dj S(k,j;t) -\int^{\kappa}_{0}dk \int_{\kappa}^{\infty} dj S(k,j;t) \nonumber \\
               & = & -\int^{\kappa}_{0}dk T^{--}(k,t) -\int^{\kappa}_{0}dk T^{-+}(k,t) \nonumber \\
               & = & 	-\int^{\kappa}_{0}dk T^{-+}(k,t),	
\end{eqnarray}
because $\int^{\kappa}_{0}dk T^{--}(k,t) =0$ by symmetry. The filtered-partitioned forms of $T(k,t)$ are readily deduced by comparison of the first and second lines of the above equation.

This is a short form of the analysis given in \cite{McComb08}, where supporting evidence base on numerical simulations is also cited. To summarise even further, if we wish to study energy transfer to higher-wavenumber modes, then we should work with $\Pi^{-+}(\kappa,t)=-\int^{\kappa}_{0}dk T^{-+}(k,t)$ rather than $\Pi(\kappa,t)=-\int^{\kappa}_{0}dk T(k,t)$, as introduced in equation (\ref{Pidef}) 

\section{The Richardson-Kolmogorov cascade}

In 1941, Kolmogorov made Richardson's earlier cascade picture the basis of a physical model of turbulence. This debt to Richardson was not stated in the these earlier works \cite{Kolmogorov41a},\cite{Kolmogorov41b} (otherwise K41) but was explicitly acknowledged in his 1962 paper \cite{Kolmogorov62} (otherwise K62). We will not repeat the analyses here, as they are well known, and a simple account can be found in, for instance, Section 4.6 of the book \cite{McComb14a}. Instead, we will confine ourselves to some quite general comments.

Although Kolmogorov actually derived the $r^{2/3}$ law for the second-order structure function $S_2(r)$, his name has always been associated with the $k^{-5/3}$ spectrum, and of course the two are trivially connected by Fourier transformation. In 1945, Onsager working in wavenumber space, argued that the neglect of both input and viscous dissipation in the inertial range implied that the inertial flux must be constant, and derived the $-5/3$ spectrum directly. He also pointed out that Fourier transformation took his result back to Kolmogorov's $2/3$ law! In a sense, two different approaches then built up, mainly in the 1960s. 

On the one hand, many measurements of the energy spectrum clearly established that at high wavenumbers spectra scaled on the Kolmogorov wavenumber and velocity, and that such spectra tended to the $k^{-5/3}$ form. There were two elements to this. The neglect of the input structure stemmed from the scale invariance of the inertial flux, whereas the neglect of the viscous term came from the NSE, with the factor of $k^2$ ensuring that the dissipation was significant at higher wavenumbers. This spectrum was seen as a suitable target for the statistical closure theories. In all, the energy spectra provided a convincing example of universal behaviour as the Reynolds number increased \cite{McComb90a},\cite{Sreenivasan95}.

On the other hand, the real-space picture became dominated by the measurement of higher-order moments and this resulted in graphs of exponent against order, which very strikingly showed a deviation from the straight line of the Kolmogorov results \cite{VanAtta70},\cite{Anselmet84}. This led to the introduction of the term \emph{anomalous exponents}, without necessarily any attribution to intermittency effects. Nevertheless workers in this field tended to be influenced by K62 \cite{Kolmogorov62}. They also tended to have various questions about his method. This is in some contrast to those working in spectral space. It perhaps reflects the fact that spectral space is a more physical way of describing turbulence, with the modes acting as degrees of freedom. Howevere, one may note that there is growing use of pseudo-spectral methods to evaluate moments or structure functions: see \cite{Qian97}-\nocite{Qian99}\nocite{Tchoufag12}\nocite{Bos12}\cite{McComb14b}.

Lastly, for completeness we should mention that there has been increasing interest in the effect of finite Reynolds numbers on the Kolmogorov theory, which assumes an infinite Reynolds number: see \cite{Qian97}-\nocite{Qian99}\nocite{Tchoufag12}\nocite{Bos12}\cite{McComb14b},\cite{Gotoh02},\cite{Antonia06}.

\subsection{Is Kolmogorov (1962) physically valid?}

We conclude this section by pointing out a number of strange features in the literature of the subject. First, the Kolmogorov theories rely on an infinite Reynolds number limit, yet many people seem to ignore the need to take account of finite Reynolds number effects in actual investigations. Secondly, in measuring higher-order moments, it is necessary to take systematic error into account. Any measured distribution will have an experimental error, which may be expected to increase towards its wings, and hence preferentially affect the higher-order moments. From this cause alone, exponents may be expected to lie below their canonical values. Yet this is never mentioned. Thirdly, and most surprisingly, while Kraichnan \cite{Kraichnan74} has pointed out that K62 is not so much a `refinement' \emph{(sic)} of K41, but in fact a radically different theory, no one seems to have noticed that it is actually unphysical. It depends on an external scale related to the physical size of the system. As the energy spectrum is an intensive quantity, it cannot depend on the size of the system in either continuum mechanics or statistical physics. Moreover, it is trivial to show that changing from K41 to K62 destroys the universal scaling of spectra which has been widely observed with the former.

\section{The viscous dissipation rate}

As is well known, the viscous dissipation rate for isotropic turbulence is given by:
\beq
\varepsilon = \nu \sum^3_{\alpha,\beta=1}\left\langle\left( \frac{\partial u_\alpha}{\partial x_\beta}\right)^2 \right\rangle,
\label{viscdiss}
\eeq
where $\nu$ is the kinematic viscosity of the fluid. Note that $\nu$ is never zero for a Newtonian fluid: we shall return to this point.

Belief in a dissipation anomaly appears to stem from a result due to Taylor \cite{Taylor35} in 1935, which used to be referred to as `Taylor's dissipation surrogate'. We quote the relevant passage from page 439 of Taylor's paper, as follows:
\begin{quote}
`rate of dissipation must be proportional, so far as changes in linear dimensions, velocity and density are concerned, to $\rho u^3/L$, where $L$ is some linear dimension defining the scale of the system.' 
\end{quote}
Note that we have changed from Taylor's original notation in that we have replaced $u'$ for the rms velocity by $u$, and $l$ by $L$, the latter denoting the integral length scale, which is invariably used in this expression nowadays. Evidently Taylor was considering how results could be scaled from one turbulent system to another but he did not dwell on the origins of this expression.

However, some people have offered simple explanations. A representative view is given by Tennekes and Lumley \cite{Tennekes72}, who refer to the amount of energy per unit mass in the large scales as being proportional to $u^2$, while the rate of transfer to smaller scales is assumed to be proportional to $u/L$. Hence we get $\varepsilon \sim u^2 \times u/L$ as the flux of energy to the smaller scales\footnote{Note the misleading use of $\varepsilon$ to represent flux in this analysis. For instance, $\varepsilon_T$ would be better, where the $T$ stands for transfer.} They commented on the `passive nature of the viscosity', in that it simply dissipates the incident energy from the large scales to heat. This, of course, is characteristic of all friction processes. 

In a more modern notation, Taylor's relation can be written as if it were the expression for the dissipation $\varepsilon$, thus:
\beq
\varepsilon = C_\varepsilon u^3/L,
\label{Tdiss}
\eeq
where the prefactor $C_\varepsilon$ may depend on the Reynolds number.  It is this expression which leads to the notion of the dissipation being independent of the viscosity, and hence being thought of anomalous. In fact, this is too hasty a conclusion, as it rules out the possibility of $C_\varepsilon$ depending on the viscosity,  which in fact it does. 

The Taylor expression for the dissipation (\ref{Tdiss}) has been widely studied in recent years, although  it is now usual to rewrite it as the dimensionless dissipation, thus:
\beq
C_{\varepsilon}(R_{L}) = \frac{\varepsilon}{U^{3}/L}.
\label{dimdiss}
\eeq
It has become clear from both experiment and numerical simulation that the dimensionless dissipation shows an asymptotic behaviour as the Reynolds number increases, such that 
\beq
C_{\varepsilon}(R_{L})\rightarrow C_{\varepsilon,\infty} \qquad \mbox{as}\qquad R_{L}\rightarrow \infty,
\eeq
where $C_{\varepsilon,\infty}$ is constant. This was highlighted in the review by Sreenivasan \cite{Sreenivasan84} and reinforced by a later review \cite{Sreenivasan98} which concentrated on the results from numerical simulations.

\subsection{The asymptotic nature of the dimensionless dissipation rate of turbulence}

More recently, an analytic form has been obtained for $C_{\varepsilon}(R_{L})$ \cite{McComb10b} and this has been verified by numerical simulation \cite{McComb15a}. Both these articles contain comprehensive citations of earlier work in the field, and reference should be made to them or to the book \cite{McComb14a} for those who wish to have a fuller picture. Here we will give a brief outline of the analysis, which is based on the supplemental material to \cite{McComb10b}. It is convenient to divide this work into separate treatments of stationary turbulence and free decay.

\subsubsection{Stationary turbulence}

We begin with the KHE for forced, stationary turbulence \cite{McComb14a}. We consider injection of energy at scales $r\geq r_{I}$ (say) and restrict attention to $r\leq r_{I}$, so that the input term does not appear in this local equation. Introducing the \emph{dimensionless structure functions} $f_{n}(x)$ by:
\beq
S_{n}(r) = U^{n}f_{n}(x)\qquad\mbox{with}\qquad x = r/L,
\label{dimstruct}
\eeq
we substitute this into the KHE and obtain for the the dimensionless dissipation, as given by (\ref{dimdiss}),
\beq
C_{\varepsilon} = A_{3}(x,R_{L}) + A_{2}(x,R_{L})/R_{L}
\eeq
where $A_{2}$ depends on $f_{2}$ and $A_{3}$ on $f_{3}$.
The two coefficients are variables, but if  we decompose them into constant and variable parts, thus:
\beq
A_{2}(x,R_{L}) = A^{c}_{2} + a_{2}(x,R_{L});
\eeq
and
\beq
A_{3}(x,R_{L}) = A^{c}_{3} + a_{3}(x,R_{L}),
\eeq
we find that:
\beq
C_{\varepsilon} = A^{c}_{2} + A^{c}_{2}/R_{L}
\eeq
where the constant $A$ is given by
\beq
A = A^{c}_{2}/A^{c}_{3},
\eeq
as
\beq
a_{3}(x,R_{L}) + a_{2}\frac{(x,R_{L)}}{R_{L}} = 0.
\eeq
Lastly, in a compact form,
\beq
C_{\varepsilon} = C_{\varepsilon,\infty} \left[L + A/R_{L}\right]
\label{finaldiss}
\eeq
where the constant $A$ is given by 
\beq
A = A^{c}_{2}/A^{c}_{3}.
\eeq

In a later treatment of forced turbulence \cite{McComb15a}, equation (\ref{finaldiss}) was derived by expanding the dimensionless structure function in inverse powers of the integral scale Reynold's number, and the coefficients fixed by fitting them to the results of a numerical simulation, thus:
\beq
C_{\varepsilon} = C_{\varepsilon,\infty} + \frac{C}{R_{L}} + O(\frac{1}{R^{2}_{L}}),
\label{ourdiss}
\eeq
with
\beq
C = 18.9 \pm 1.3\quad\mbox{and}\quad C_{\varepsilon,\infty} = 0.468 \pm 0.006.
\label{ourconstants}
\eeq
This is illustrated in Figure 1, which is reproduced from \cite{McComb15a}.

A subsequent application to magnetohydrodynamics required the second-order term to be taken into account, although the effect was small \cite{Linkmann15a}.

\begin{figure} 
\begin{center}
\includegraphics[width=0.75\textwidth, trim=0px 0px 0px 0px,clip]{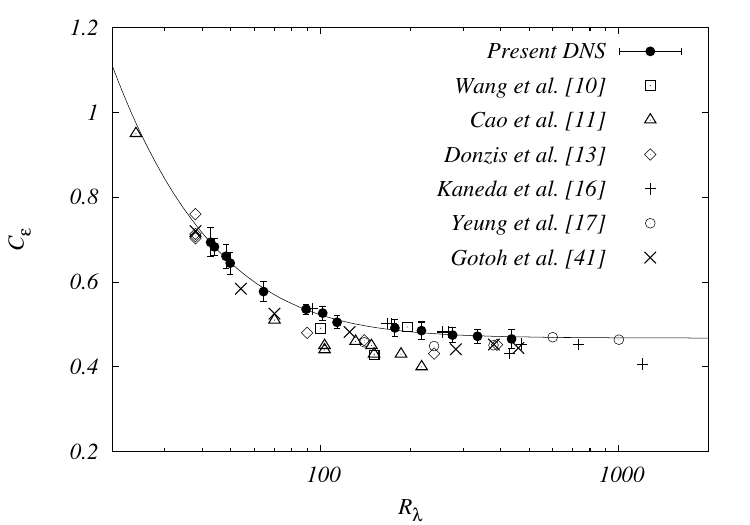}
\end{center} 
\caption{\small This figure shows the dimensionless dissipation rate plotted against Taylor-Reynolds number and is reproduced from the article \cite{McComb15a} The continuous line  labelled as `Present DNS' is a plot of equation (\ref{ourdiss}) with constants given by equation (\ref{ourconstants}). The citations to other investigations may be obtained by reference to the article \cite{McComb15a}}. 
\label{fig1} 
\end{figure}

\subsubsection{Freely decaying turbulence}

In free decay we introduce time dependence into all statistical quantities so that we have $\varepsilon(t)$ and $S_{n}(r,t)$ at any time $t$. In order to make comparisons, we introduce a fiducial time $t = t_{e}$, when the turbulence has evolved to a state which depends on the NSE only. As above, we begin with the change of variables, and introduce dimensionless structure functions $g_n(x,\tilde{t})$,
\beq
S_{n}(r,t) = U^{n}(t_{e})g_{n}(x,\tilde{t}),
\eeq
where
\beq
x = r/L(t_{e});\quad \tilde{t} = t/T;\quad T = L(t_{e})/U(t_{e}).
\eeq
The time dependence leads to an additional coefficient
\beq
B_{2} =\left. \frac{3}{4} \frac{\dd g_{2}}{\dd t}\right |_{t_{e}},
\eeq
and hence
\beq
C_{\varepsilon}(t_{e}) = C_{\varepsilon,\infty}\left[1 + \frac{1}{R_{L}} \frac{A^{c}_{2}}{A^{c}_{3} - B^{c}_{2}}\right]
\eeq

This general picture was well supported by the results of numerical simulation \cite{McComb10b}. It was also printed out that the presence of the term $\partial S_{2}/\partial t$ means that Onsager's criterion for the existence of an inertial range (i.e. that the peak inertial flux becomes equal to the dissipation) can never be completely satisfied. This had previously been noted by Sagaut and Cambon in the book \cite{Sagaut08}.

\subsection{Turbulent dissipation is not anomalous}

Equation (\ref{Tdiss}) is written as if it defines the dissipation rate, but in fact it only asserts that the quantity on the rhs is equal to the dissipation. In other words, the flux from larger scales is equal to the dissipation in small scales.

In order to emphasise this point, let us consider the very simple example of laminar Poisieuille flow in the $x$-direction between parallel plates situated as $y=\pm h$. As is well known, the velocity $u(y)$ is given by:
\beq
u=\frac{P}{2\mu}\left(h^2-y^2 \right)=\frac{3U}{2h^2}\left(h^2-y^2 \right),
\eeq
where $P$ is the constant pressure gradient, $\mu$ is the dynamic viscosity, and $U$ is the bulk mean velocity. The latter quantity is defined in terms of the volumetric flow rate $Q$. As is also well known, the viscous dissipation per unit area is given by:
\beq
\varepsilon=\int_{-h}^h \mu \left(-\frac{3U}{h^2}y\right)^2 dy= \frac{6\mu U^2}{h}.
\label{diss}
\eeq

Now the fluid flows because of the action of the pressure, and the rate of doing work by the pressure forces is $QP$. For steady flows, this must be equal to the rate of viscous dissipation, or:
\beq
QP = \varepsilon.
\label{work}
\eeq
However, writing this equation the other way round, as is so often done with Taylor's expression, we have:
\beq
\varepsilon = QP.
\label{fakediss}
\eeq
But this does not define the dissipation and hence does not lead us to the false conclusion that the dissipation in Poiseuille flow does not depend on the viscosity. In fact we know that it does, when we look at equation (\ref{diss}). Exactly the same rebuttal applies to literally every turbulent flow: we may write down an exact expression for the viscous dissipation rate in terms of the fluid viscosity. In the case of isotropic turbulence, this is just equation (\ref{viscdiss}).

\subsection{The  infinite Reynolds number limits of isotropic turbulence}

In recent years there has been growing interest in the idea of an infinite Reynolds number limit; but, at the same time, there has been quite a lot of confusion about what this should mean, including cases where the physical significance of the limit is unclear and it may not be a limit at all. One source of confusion can arise due to the fact that the Navier-Stokes equation can be derived as a part of Continuum Mechanics or in the continuum limit of statistical mechanics. Although the two forms of the equation are superficially identical, their different derivations may have implications for the taking of limits. It is for this reason that we may have to consider more than one form of infinite Reynolds number limit.

\subsubsection{The infinite Reynolds number limit in Continuum Mechanics}

The first instance of a formal limit being taken was in 1953, in the first edition of Batchelor's book \cite{Batchelor71}, in which he considered the limit: $\lim \nu \to 0$ at constant dissipation rate $\varepsilon$. He took the Kolmogorov dissipation wavenumber $k_d$ as a measure of the largest wavenumber present, and hence:
\[
\lim_{\nu \to 0}  k_d   = \lim_{\nu \to 0} (\varepsilon/\nu^3)^{1/4} = \infty, \, \mbox{for}\, \varepsilon=  \mbox{constant}.
\]
Thus, as we decrease the viscosity, the volume of the wavenumber space increases, and the system can absorb ever more energy. 

Note that it would really be more physical if we considered a forced NSE, and carried out the limiting procedure at constant rate of input $\varepsilon_W$ (say). This was implicit in the procedure followed by Edwards in testing his SCF closure theory \cite{Edwards65}. He argued that in the limit the dissipation term in the energy balance equation could be represented by a delta function, thus: $\varepsilon \delta (k-\infty$; and the input localised at the origin, thus: $\varepsilon \delta (k)$.

In the context of continuum mechanics, the fluid continuum is a model for a real fluid, and we may, if we wish, take that continuum to be infinitely divisible. Accordingly, Edwards could make such an assumption in order to test his statistical closure in the limit of zero viscosity. Of course we cannot actually take the viscosity to be zero in a real Newtonian fluid. While this is a typical manoeuvre in physics; for instance, the continuum limit in quantum physics amounts to taking Planck's constant equal to zero, we know that Planck's constant is a universal constant, and is never zero. It is assumed to be so small compared to some other quantity that it can be neglected. The same must also be true of the viscosity and, in passing, it is worth noting that, when taking limits, one should not be dealing with quantities that possess dimensions. We could replace the viscosity by a non-dimensional normalised viscosity, and an obvious choice would be $\nu' = 1/R$, where $R$ is some Reynolds number. If we wish to increase the Reynolds number, then in terms of physically realizable experiments, we would put energy into the system at a higher rate, rather than decrease the viscosity. This necessitates our considering a real physical fluid and in physics the NSE for a gas is derived from statistical mechanics. Hence, a breakdown of the continuum approximation could happen when the wavenumber is large enough to be comparable to the reciprocal of the mean free path as a measure of the average distance between molecules.

\subsubsection{The infinite Reynolds number limit in macroscopic Physics}

Later on in his book, after discussing the case of infinite Reynolds number limit, Batchelor \cite{Batchelor71} gave a brief discussion of the validity of the validity of the continuum approximation, and concluded that it is satisfactory for typical turbulent flows. There is, of course, an inconsistency there, in that his infinite Reynolds number limit leads to case that could not be satisfied by the mean molecular free path. More specific calculations were given by Leslie \cite{Leslie73}, who considered a flow in a pipe of $10$mm diameter at a Reynolds number of $10^6$ (which he described as very extreme conditions) and calculated that it would have a minimum eddy size of $10^{-4}$mm. He pointed out that the mean free path in a liquid would be of the order of atomic dimensions at $10^-7$mm so that the continuum limit would be satisfied. He added that this would still be true for a gas which was sufficiently dense to produce that particular Reynolds number in such a narrow pipe, although not by so comfortable a margin.

A more extensive discussion can be found in Section 1.5 of the book by Tennekes and Lumley \cite{Tennekes72}, where they conclude that for some astronomical examples it may be necessary to resort to statistical mechanics where the continuum limit is no longer valid. More recently, Moser \cite{Moser06} has examined the validity of the Navier-Stokes equation at high Reynolds numbers and concluded that even in the limit of infinite Reynolds number, the NSE is the relevant fluid model.

\subsubsection{The case $\nu=0$ is not a limit}

In recent years there has been a certain amount of confusion about the concept of the infiniite Reynolds number limit, with statements to the effect that it is equivalent to the turbulence being governed by the Euler equation; or that there are discontinuities limiting the process of the energy cascade. These seem to stem from the interest that some mathematicians have in proving the \emph{Onsager conjecture}, which is to the effect that the Euler equation becomes dissipative under circumstances where the continuum limit breaks down. Bearing in mind that, in the absence of viscosity, there can be no dissipation in the normal sense of fluid dynamics. That is to say, viscous dissipation, in which the macroscopic kinetic energy is changed into thermal energy (i.e. the kinetic energy of molecular motion). Instead, the nonlinear term is supposed to become non-conservative due to a loss of smoothness in the Fourier representation. The relation of any of this to the nature of actual physical breakdown of the macroscopic continuum is unclear.

In order to show this, it is necessary for the continuum limit to remain valid for the local time derivative while the nonlinear term (the convective derivative) becomes non-conservative. In that sense the inertial transfer term presumably acts as `the canary in the coalmine' for the breakdown of continuity. One can see that this might well be the case, but it is a purely mathematical point and of no physical relevance. Whereas, in the case of a real physical breakdown, there is no reason to believe that it would take this form.

Onsager put forward this conjecture in 1949 \cite{Onsager49} on the following grounds. First, from Taylor's expression for the dissipation, as given in modern form in (\ref{Tdiss}), he concluded that the dissipation was  inviscid. But this is not an expression for the dissipation: it is an expression for the inertial transfer, which becomes equal to the dissipation in the large Reynolds number limit. We have discussed this in Section 5.1 above. Secondly, he took the case $\nu=0$ as the infinite Reynolds number limit. Apart from the fact that, as we have already stated, this cannot be true of any real fluid, it is also a bizarre way of taking a limit.

As an example, let us consider the function $f(\nu)= \sin \nu/\nu$, which is part of the DNA of every physicist.  If we simply set $\nu=0$, then we get the indeterminate form: $f(0)=0/0$. We have not taken a limit, which is a process. Here, our process of taking a limit can be to use l'h\^{o}pital's rule; or expand $\sin \nu$ as a Taylor series in powers of $\nu$; or to simply draw the graph of the function. The result is of course that the correct limit is $\lim_{\nu \to 0} f(\nu)= 1$.

Interest in Onsager's conjecture was stimulated by the review article of Eyink and Sreenivasan \cite{Eyink06}. This dealt with the work of Lars Onsager and contains details of his notes and correspondence with his contemporaries. Section IV of this article is a self-contained review of \cite{Onsager49}, and also gives references to the mathematical work stimulated by the conjecture.

\section{Two-time correlations and temporal spectra}

A significant unresolved issue in turbulence theory is the nature of the two-time correlation and the role of convective sweeping in its decorrelation. An analysis was given in the book by Tennekes and Lumley in 1972 \cite{Tennekes72} which presented a picture in accordance with the Kolmogorov theory. Then, in 1975,Tennekes \cite{Tennekes75} came to quite a different conclusion. Here we will give a brief account of these theories and then point to certain aspects which seem to merit further attention.

We begin with the general two-point, two-time correlation tensor $C_{\alpha\beta}(\mathbf{x},\mathbf{x'};t,t')$, as defined by equation (\ref{gen_corr}) where $\alpha$ and $\beta$ are the cartesian tensor indices, taking the values $1,2$ or $3$, and we work with the modified form which is given by equation (\ref{mod_gen_corr}) in terms of sum and difference variables. In everything that follows we will restrict our attention to homogeneous turbulence and consider a fixed point in space. This means that we may simplify the notation by omitting the space variables.  We also restrict our attention to isotropic turbulence, which means that we can replace the correlation tensor by a single scalar correlation function which we will denote by $C_E$, where the subscript '$E$' denotes `Eulerian'. Thus, for isotropic turbulence,
\begin{equation} 
C_{\alpha\beta}(\mathcal{T},\tau) \rightarrow C_E(\mathcal{T},\tau) 
\end{equation} 
Later we will introduce the Lagrangian correlation function.

\subsection{The analysis of Tennekes and Lumley (1972)}

Now, at this stage, we have imposed all the restrictions that Tennekes and Lumley made in specifying their problem. However their subsequent analysis seems to imply that they are also considering stationary turbulence, and this is an important point. We will underline this fact by continuing to treat the problem more generally, and retaining the 'clock time' $\mathcal{T}$. 

The energy spectrum  $\phi_E(\mathcal{T},\omega)$ is defined by the Fourier transform, \begin{equation}
C_{E}(\mathcal{T},\tau) = \int_{-\infty}^\infty \exp(i\omega \tau) \phi_E(\mathcal{T},\omega)d\omega,
\end{equation}
where $\omega$ is the angular frequency; and the Fourier pair is completed by: 
\begin{equation}\phi_E(\mathcal{T},\omega)= \frac{1}{2\pi}\int_{-\infty}^\infty \exp(-i\omega \tau)C_E(\mathcal{T},\tau)d\tau.
\end{equation}

As a preliminary to considering the inertial-range form of $\phi_E(\mathcal{T},\omega)$ we need to establish its dimensions. If we integrate the spectrum over all frequencies, we have: \begin{equation} \int_{-\infty}^{\infty}\phi_E(\mathcal{T},\omega)d\omega = U^2(\mathcal{T}),\end{equation} where $U$ is the root mean square velocity. Recall that $\mathcal{T}$ is the `clock time', as opposed to the `difference time' $\tau$. From this relationship it follows that the dimensions of the spectrum are: 
\begin{equation} [\phi_E(\mathcal{T},\omega)] = L^2 T^{-1}, 
\end{equation}
 where, as usual, square brackets indicate the dimensions of a quantity.

At this point we assume stationarity, which is in effect what Tennekes and Lumley did \cite{Tennekes72}, and so we omit the dependence on $\mathcal{T}$. Having, \emph{in effect}, done this, they applied the well known argument of  Kolmogorov to limit the dependence of the spectrum to the two independent variables $\omega$ and the dissipation rate $\varepsilon$. They stated that the only dimensionally consistent result is: 
\begin{equation}
 \phi_E(\omega) \equiv f(\varepsilon, \omega) = \beta \varepsilon \omega^{-2},
 \end{equation} 
 where $f$ is some arbitrary function, \emph{assumed to be a power}, and $\beta$ is a constant. Checking the dimensions, we find:
\begin{equation} [\phi_E(\omega)] = (L^2 T^{-3})T^{2} =  L^2 T^{-1},
 \end{equation}
  as required.

Later Tennekes presented a different analysis \cite{Tennekes75} in which he argued that the inertial-range temporal spectrum would be determined by convective sweeping and and this led to the result: \begin{equation}\phi_E(\omega)= \beta_E \varepsilon^{2/3}U^{2/3}\omega^{-5/3}. \end{equation} It is readily verified that this result has the correct dimensions, thus: \begin{equation} [\phi_E(\omega)] = (L^2T^{-1})^{2/3}(LT^{-1})^{2/3}T^{5/3}= L^2T^{-1}.\end{equation}

It should be noted that irrespective of the merits or otherwise of this analysis by Tennekes, it is limited to stationary turbulence \emph{in principle} due to omission of any dependence on the clock time $\mathcal{T}$. 

\subsubsection{Temporal spectra: the Lagrangian case}

Some may take the view that the discussion of the Lagrangian case along with the Eulerian case, which is the one that is of more practical importance, is an unnecessary complication. At the same time, one has to acknowledge that the application of these ideas to the assessment of statistical closure theories should take account of the fact that there are Lagrangian theories as well as Eulerian theories. Also, there is an interesting general point to be made when we compare the treatment in the book by Tennekes and Lumley \cite{Tennekes72} with the later analysis of Tennekes \cite{Tennekes75}.

So far we have only mentioned the discussion by Tennekes and Lumley  of the inertial range behaviour of the Eulerian spectrum. In fact they not only derived the inertial range form of the Lagrangian spectrum, and found it to be the same power law as the Eulerian case, but also obtained a relationship between the constants of proportionality in the two cases.

The crucial step in this work is the equivalence of the two correlations (see Section 8.5 of \cite{Tennekes72}), where the authors refer back to their discussion of Lagrangian forms in Section 7.1 \emph{ibid} (actually they incorrectly give this as 7.2). Following their notation, we represent the Lagrangian velocity of a fluid point by $V_{\alpha}(t)$ where $\alpha = 1, 2, \mbox{or}, 3$. Then, they assert that $\langle V_{\alpha}V_{\alpha}\rangle =\langle u_{\alpha} u_{\alpha}\rangle$, where $u_{\alpha}$ is of course the Eulerian velocity; leading on to their equation (8.5.3). This is the step that provides the basis for their assertion of the equivalence of the Eulerian and Lagrangian inertial range spectra.

However, the later work of Tennekes \cite{Tennekes75} led to the Eulerian spectrum being different from the Lagrangian form, due to the supposed predominance of sweeping effects. This would seem to give rise to an inconsistency, and we will return to this  when we examine the work of Tennekes more closely.

Eearlier, we noted that the form of two-time correlation being studied in \cite{Tennekes72} was limited to stationary flows. This point was also made by Hinze \cite{Hinze59}: see equation (1-57), page 39 in the first edition. However, in discussing the motion of fluid points in Lagrangian coordinates, Tennekes and Lumley emphasised the need for both homogeneity and stationarity. So in effect this restriction had already been made. We also note that an alternative discussion of the original work by Lumley can be found in Section 12.2 of \cite{McComb90a}.

\subsection{The analysis by Tennekes (1975)}

We now take a closer look at the analysis by Tennekes, in which he differed from his earlier analysis with Lumley, and concluded that large-scale sweeping is the determining factor in the decorrelation of the two-time correlation in the inertial range. As noted earlier, this leads (rather confusingly) to a `$-5/3$' power law for the Eulerian temporal spectrum, when the Kolmogorov form is actually $n=-2$. His starting point is equation (1) in \cite{Tennekes75}, which may be written in our present notation as: 
\begin{equation}
\frac{\partial u_1}{\partial t}=-\left(u_1\frac{\partial u_1}{\partial x_1}+u_2\frac{\partial u_1}{\partial x_2}+u_3\frac{\partial u_1}{\partial x_3}\right),
\label{Ten_Taylor}
\end{equation} 
and this was justified by his assuming that Taylor's hypothesis of frozen convection applies.

The usual application of Taylor's hyopothesis is to situations where there is a mean or free stream velocity $U_1$,  which is much larger than the turbulent velocity $\mathbf{u}(\mathbf{x},t)$. Then the changes in the velocity field with time at a fixed measuring point could be due to the passage of a frozen pattern of turbulent motion past that point. Hence the local time derivative at a point may be replaced by  the convective derivative, thus: 
\begin{equation}
\frac{\partial}{\partial t} \rightarrow -U_1\frac{\partial}{\partial x_1} \quad \mbox{if} \quad U_1 \gg u.
\end{equation}
Or in the context of spectra,
\begin{equation}k_1 = \omega/U_1.
\end{equation}
A fuller discussion of this can be found in Section 2.6.5 of the book \cite{McComb90a}.

Thus equation (\ref{Ten_Taylor}) seems a rather extreme application of Taylor's hypothesis. In fact we can write down an exact expression for ${\partial u_1}/{\partial t}$ by invoking the Navier-Stokes equation. This gives us
\begin{equation}
\frac{\partial u_1}{\partial t}=-\left(u_1\frac{\partial u_1}{\partial x_1}+u_2\frac{\partial u_1}{\partial x_2}+u_3\frac{\partial u_1}{\partial x_3}\right)-\frac{\partial p}{\partial x_1} + \nu \nabla^2 u_1,
\end{equation} 
where $p$ is the kinematic pressure and $\nu$ is the kinematic viscosity. Thus in using equation (\ref{Ten_Taylor}), Tennekes neglects both the pressure and the viscous terms. The latter may seem reasonable, as his main concern was with the inertial range, but it must be borne in mind that the subsequent analysis involves squaring and averaging both sides of equation (\ref{Ten_Taylor}), so the neglect of the viscous term may introduce significant error. However, the neglect of the pressure term is even more concerning, as this is a highly non-local term with the pressure being expressed in terms of integrals of functions of the velocity field over the entire system volume: see Section 2.1 of \cite{McComb90a}.

This analysis by Tennekes relies on imponderable assumptions about scale separation and statistical independence. Such ideas were to be discussed much later on, and rather more quantitatively, in the context of mode elimination and large-eddy simulation: see Chapter 8 in the book \cite{McComb14a} for an account of this work. It is clear that the analysis by Tennekes has swept a great deal under the carpet. In contrast, the arguments given by Tennekes and Lumley \cite{Tennekes72} seem, to me at least, more confident and well justified than those given in \cite{Tennekes72}. In his conclusion, Tennekes remarked on the difference between the two analyses, stating that it was `embarrassing in a personal sense.' Certainly both sets of arguments could repay closer study.

As a final point, Tennekes expressed the view that the implications of his later work \cite{Tennekes75} supported Kraichnan's view that Lagrangian coordinates are better suited to statistical closure theories than the more usual Eulerian variety. However, it is worth pointing out that all the analyses that support such a view are valid (if at all) only for stationary turbulence, whereas virtually all the numerical assessments of closure theories are restricted to freely decaying turbulence. We shall return to this point later on, when we discuss statistical theories.

\section{Renormalization and field-theoretic approaches}

Here we come to the most difficult and fraught aspect of turbulence theory, in that we are faced with nothing less than a clash of cultures. The study of turbulence, from the nineteenth century on, had always been dominated by engineers and applied mathematicians, while the more recent renormalization methods reflected the post-war development of quantum field theory and are carried out by theoretical physicists. It is worth remarking that the pioneers of this activity had a clear understanding of the similarities, and more importantly the differences, between quantum field theory and the theory of turbulence. This was also true of the later pioneers of renormalization group (RG) applied to fluid motion, reflecting Wilson's successful application of field-theoretic RG to the study of critical phenomena. Unfortunately, existing turbulence researchers were not well placed to understand these fine distinctions, nor to appreciate that many later effusions from theoretical physicists over the years have concealed their lack of relevance to turbulence behind a great deal of impenetrable mathematics, buttressed by Feynmann diagrams (which have no relevance to turbulence although, as we shall see, Wyld diagrams do) and various sweeping and glib remarks which justify nothing. It is my aim here, to help the interested reader to separate the wheat from the chaff, and to understand what has been achieved that is almost certainly of permanent value.

\subsection{What is Renormalization?}

Let us begin by demystifying the word `renormalization'. The concept, although not the word, began right here in turbulence research in the late nineteenth/early twentieth century, with the introduction of the effective turbulent diffusivity and viscosity. The idea was that a molecular coefficient, originating in the random motion of molecules, could be supplemented by the random motions of macroscopic turbulent eddies. In 1923, the Debye-H\"{u}ckel theory of an electolyte introduced the idea of a the \emph{screened potential}. However, this could also be interpreted in terms of a Coulomb potential of a renormalized charge which was scale dependent. With the development of quantum electrodynamics in the 1940s, leading on to quantum field theory, the need for renormalization arose. In quantum mechanics, a particle is represented by a wave function which must be normalized. When a bare particle is `dressed' by interactions, the resulting \emph{quasi-particle} must have its wave function \emph{renormalized}. A very simple and accessible introduction to these concepts, and how they may be applied to turbulence, may be found in the book \cite{McComb04}. Here we begin by dividing the various approaches to turbulence into three categories.

\subsection{The three main types of renormalization theory as applied to turbulence}

When we consider the NSE in real ($x$) space\footnote{Note that by `real' we mean in the everyday sense of the real world we live in, rather than as opposed to `imaginary', although we should note that the Fourier-transformed velocities are complex in the ordinary mathematical sense.}, we are confronted by a nonlinear partial differential equation for which there is no general method of solution and which is not amenable to perturbation theory except for the case of creeping flow, where the Reynolds number is very small. On the other hand, if we Fourier transform the NSE into wavenumber ($k$) space, the problem becomes one of many-body statistical physics, where the velocities in wavenumber space are the degrees of freedom. The resemblance of this to quantum-field theory was noted in the 1950s and in 1966, Edwards commented that turbulence was a problem in field theory where the coupling constant (i.e. the Reynolds number) could be varied in the laboratory from zero up to infinity.

We discuss these renormalization approaches chronologically, based on the date of the first significant published contribution in each type of theory. 

\subsubsection{Statistical closures using renormalized perturbation theory}

The first attempt at a fundamental statistical closure was the quasi-normality hypothesis in the early 1950s. As is well known, this failed due to its unphysical prediction of negative energy spectra. We will not discuss it here and refer the reader to Section 3.4 of the book by Leslie \cite{Leslie73} or Section 2.8.2 of \cite{McComb90a}. Its main effect was to lead those who followed to place considerable emphasis on \emph{realizability} of theories. Kraichnan began publishing on turbulence in the late 1950s, culminating in his direct-interaction approximation (DIA) \cite{Kraichnan59b}.

His theory was for the spectral form of the two-point, two-time covariance, as denoted for isotropic turbulence by equation (\ref{spec_density}), and began by introducing the infinitesimal response tensor $\widehat{R}^{DIA}_{\alpha\beta}\mathbf{k};t,t')$, which is defined by:
\beq
\delta u_{\alpha}(\mathbf{k},t)= \int \widehat{R}^{DIA}_{\alpha\beta}(\mathbf{k};t,t') \delta f_{\beta}(\mathbf{k},t')dt',
\label{inf_resp}
\eeq
where $\delta f_{\beta}(\mathbf{k},t')$ is a fluctuation in the random stirring forces which are introduced to generate the turbulence, $\delta u_{\alpha}(\mathbf{k},t)$ is the resulting fluctuation in the random velocity field, while the hat symbol indicates that the response tensor is a random variable. He obtained an equation for the response function by functional differentiation of the NSE which could be solved simultaneously with the equation for $C(k;t,t')$, which was obtained in the usual way from the NSE. Both equations were solved by perturbation theory, using an expansion in an ordering parameter $\lambda$ which was put equal to unity at the end of the calculation. These series were, of course, wildly divergent; and renormalization was achieved by summing classes of terms to all orders. 

In the course of an unusual version of perturbation theory, Kraichnan assumed that the response tensor was statistically independent of the velocity field, such that (for any wavenumber):
\beq
\langle \widehat{R}^{DIA}(t,t')u(t)u(t') \rangle = \langle  \widehat{R}^{DIA}(t,t')\rangle\langle u(t)u(t') \rangle = R^{DIA}(t,t')\langle u(t)u(t') \rangle,
\eeq
where the ensemble-averaged response function is introduced, and for anisotropic turbulence, is a function of the wavenumber and two times only. Thus, the theory yields two simultaneous expansions for $C(k;t,t')$ and $R^{DIA}(k;t,t')$, which are truncated at lowest nontrivial order. The replacement of the instantaneous response function by the ensemble-averaged form is a type of mean-field theory, so it is worth remarking that the same results were later found by conventional perturbation theory, thus justifying the mean-field step in the DIA.

\subsubsection{Statistical formalisms}

The difference between a statistical formalism and a theory, is that the formalism is trying to establish the existence and nature of a general solution to the closure problem. There are two significant attempts to do this: the diagrammatic method of Wyld \cite{Wyld61} and the functional formalism of Martin, Siggia and Rose \cite{Martin73} (often referrred to as MSR, for short). 

First, in 1961 Wyld published a formal perturbation treatment of the turbulence problem, by using diagrams to represent the terms in the perturbation series and to show how renormalization may be achieved by resummation \cite{Wyld61}. This is a very attractive way of introducing renormalization in turbulence, with the topology of the diagrams playing a part, although they do not have the physical significance that the Feynmann diagrams have in particle theory. However, unfortunately, the Wyld formalism, as initially presented, is incorrect. This was first signalled by Lee in 1965, who noted that with Wyld's method there was a problem of double counting of diagrams which could be cured by an \emph{ad hoc} modification \cite{Lee65}. Later on, MSR \cite{Martin73} claimed that the representation of vertex terms in the Wyld theorywas incorrect, while in 1978 McComb \cite{McComb78} showed that the double-counting problem arose from a procedural error, which was easily rectified. More recently, Berera \emph{et al.} \cite{Berera13} analysed both the formalisms and argued that they were compatible if one introduce an \emph{Improved Wyld-Lee Renormalized Perturbation Theory}, which embodied the changes originally suggested by McComb \cite{McComb78}.

In 1973, MSR published a much more ambitious attempt at introducing a functional formalism, by emulating the way in which the Hamilton-Lagrange formalism of classical mechanics had been extended to quantum field theory \cite{Martin73}. As turbulence is not a Hamiltonian system, this involved the synthesis of a quantity to play the part of the action. In doing this they relied on the introduction of stirring forces to generate an adjoint field; a step which was equivalent to Kraichnan's \emph{ansatz} for the DIA. This theory attracted a lot of attention but is notoriously difficult to understand. For those who still wish to understand MSR, some degree of exegesis can be found in \cite{Berera13}.

To summarise, we can say the following about these two formalisms:
\begin{enumerate}
\item Both formalisms consist of simultaneous renormalized expansions for the covariance and renormalized response functions.
\item Both formalisms are cognate to Kraichnan's DIA in that the response is defined in terms relating fluctuations in the velocity field to fluctuations in the stirring forces.
\item Both formalisms are identical to Kraichnan's DIA when truncated at lowest nontrivial order.
\item Both formalisms have rrecently been shown to be equivalent to each other, as indeed they must be, because  ultimately all functional formalisms are justified by reference to perturbation theory.
\end{enumerate}
We shall return to these points later when discussing the failure of the pioneering theories.

\subsubsection{Renormalization Group (RG)}

Wilson's application of field-theoretic RG to critical phenomena \cite{Wilson75a} stimulated great interest in the method and effectively founded the subject of \emph{Statistical Field Theory} \cite{Lebellac91}. Its success lay not just in the fact that it provided a powerful method of calculation, but also in its ability to expose the underlying physics of critical phenomena. As we shall see, this is also true of its application to turbulence, but sadly this fact has not penetrated the turbulence community to a similar extent. My aim here is to explain this point of view, and we begin with a very brief consideration of the behaviour of a magnetic lattice.

We may envisage a ferromagnet as a regular three-dimensional lattice, with a molecular magnet at each lattice site. The case of interest is when all the molecular fields line up with each other at a finite temperature $T\leq T_C$ where $T_C$ is referred to as the \emph{Curie temperature}; or, more generically, the \emph{critical temperature}. In this case, RG is a coarse-graining technique, in which one averages over a number of unit cells to construct a new lattice and then re-scales to take it back to the previous lattice. When the current and previous lattice are the same, the procedure has reached a \emph{fixed point}. This is a form of \emph{scale invariance}.

The molecular magnets have a tendency to align, and it is only thermal agitation that prevents this happening. Thus it follows that at $T=0$ all magnets are aligned and this situation will be invariant under RG transformations and is a fixed point. Similarly, at very high temperatures, spins will be oriented at random, and once again the situation is invariant under RG transformation and is at a fixed point. These are referred to as \emph{trivial fixed points}. As the temperature is lowered towards the critical temperature, fluctuations occur in which spins line up over increasing distances. When the correlation length of these fluctuations tends to infinity, RG transformations once again reach a fixed point, but this time at a temperature $T=T_C$, and this is known as the \emph{critical fixed point}.

This is a rather picturesque treatment of what is known as \emph{real space RG} and in reality the RG transformations are applied to the Hamiltonian of the system. A simple but much fuller introduction to this topic can be found in \cite{McComb04}. Noting that fluid turbulence does not possess a Hamiltonian, we now have to consider how to approach the NSE. The first question is: can we employ a similar simple physical argument to the turbulent case? As the magnetic case is an example of a phase transition, this might seem more appropriate if applied to the laminar-turbulent transition, but in fact we will be concerned with developed turbulence. Accordingly, we consider stationary turbulence, driven by random (Gaussian) forces at low wavenumbers and characterised by a fluid viscosity $\nu_0$; a notation indicating that we expect to renormalize this viscosity. We will be operating in wavenumber space, which means that we will be using the inverse lattice, and that RG transformations will involve reducing the wavenumber.

It is well known that the analogue of the coupling constant for turbulence is the Reynolds number, and we shall follow Batchelor \cite{Batchelor71} and assign a local Reynolds number $R(k)$ to each individual degree of freedom, thus:
\beq
R(k)=[E(k)]^{1/2}/\nu_0 k^{1/2}.
\label{local_re}
\eeq
The value of this coupling constant therefore determines the difficulty of solving the NSE at any particular value of the wavenumber $k$. In Figure 2, we show a schematic of the turbulent energy spectrum as a function of wavenumber, and from this we can make some deductions about the local coupling constant as given by (\ref{local_re}). 

\begin{description}
\item[Low wavenumber limit] As $k \rightarrow 0$, it is well known that we may write the turbulent energy spectrum as:
\beq
E(k)=E_2 k^2 + E_4 k^4 + \mathcal{O}(k^6).
\eeq
Recently it has been shown that $E_2 =0$ \cite{McComb16a}, and so the low-wavenumber spectrum varies as $k^4$ as $k \rightarrow 0$. From (\ref{local_re}), it follows that the local coupling constant vanishes in that limit. In field-theoretic terms, this is asymptotic freedom in the infra-red. As the coupling tends to zero, the velocity field is determined by the stirring force and hence is Gaussian in nature. This means that a fixed point at $k=0$ is analogous to a high-temperature fixed point in the magnetic case, as it corresponds to complete disorder: if it exists, this is a trivial fixed point.
\item[High wavenumber limit] Experimental results taken over many decades have indicated that at high wavenumbers the spectrum falls off faster than a power and is generally taken to be some form of exponential decay. Accordingly, from (\ref{local_re}) it is clear that the local coupling must fall off exponentially with increasing wavenumber. In this region, viscous effects dominate and the turbulence tends to a more ordered state. Hence, if there is a fixed point as $k \rightarrow \infty$, it will be analogous to the trivial low-temperature fixed point in the magnetic case. It also corresponds to asymptotic freedom in the ultra-violet.
\end{description}

This leaves us with the possibility of a non-trivial fixed point corresponding to the top of the inertial range, as RG transformations decrease the wavenumber from the dissipation range. This turns out to be the case and we will discuss that in a later section.

\begin{figure} 
\begin{center}
\includegraphics[width=0.65\textwidth, trim=0px 0px 0px 100px,clip]{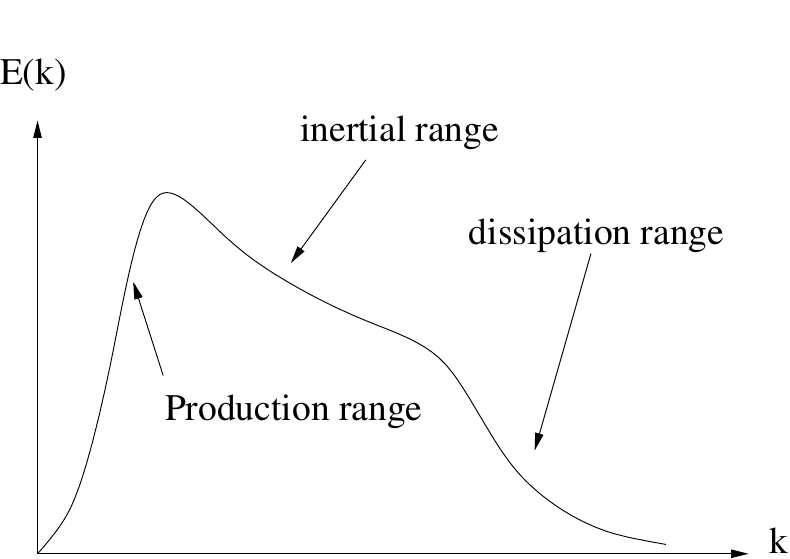}
\end{center} 
\caption{\small  Schematic view of the energy spectrum for turbulence at a Reynolds number large enough to exhibit an inertial range.} 
\label{fig2} 
\end{figure}

\subsection{Two-point statistical closures}

In the last century, the statistical theory of turbulence was dominated by the work of Kraichnan; not just by reason of his pioneering DIA, but by his prodigious output of ideas, particularly in his analysis of the failure (as he saw it) of the DIA, and its extension to Lagrangian coordinates. So, to some extent, this means that other works have been overlooked, not least the Local Energy Transfer (LET) theory, which contains within it a diagnosis of what went wrong with DIA. We begin by discussing the rival theories, how they are related to each other, and how they perform in numerical calculations.

In 1964, Edwards published his self-consistent field theory of isotropic turbulence \cite{Edwards64}, in which he derived a form of the Liouville equation appropriate to the turbulence problem. He saw the advantage of this as its linearity in the probability distribution, and the possibility of obtaining a correction to mean-field theory by calculating corrections in terms of the deviation of the exact distribution from a base-model Gaussian. He restricted his theory to the single-time case and introduced stirring forces that generated white noise, as is usual in dynamical systems theory. In doing this he was guided by statistical treatments of Brownian motion, and this is something which we should keep in mind. Later Kraichnan showed that, if one assumed exponential time-dependences in the DIA, the resulting equations were similar to those of Edwards. In fact, the equation for the energy spectral density was identical, while the equation for the response function contained only two lifetimes as compared to three in the Edwards form. This is a point that we should highlight:
\begin{quote}
The Edwards theory, and the Kraichnan theory, lead to the same equation for the energy spectral density, despite their very different kinds of derivation.
\end{quote}
As a corollary, we might add: all theories of turbulence lead to the same energy equation and, if they differ, it is in the  form of the response function equation.

In 1965, Herring's very elegant self-consistent field method appeared and this was extended to the two-time case in the following year. A recent accessible treatment of this theory may be found in the commemorative issue of this journal \cite{McComb23A}. The energy balance equation for the single-time case was identical to  that  of Edwards, while the two-time form was thought to be equivalent to the DIA.

All three theories were found to be incompatible with the Kolmogorov spectrum and there was much discussion about this at the time. Generally it was agreed that the problem lay in the response equation and was something to do with its behaviour at low wavenumbers; but it was not until 1974 that the present author realised that the relationship between the response equation and the energy equation in the Edwards theory (and by extension, the other two) was not in accord with the more recently measured energy transfer spectrum in isotropic turbulence \cite{McComb74}. This led to the concept of the turbulent response being determined by the complete transfer spectrum $T(k)$; and not, as in the three pioneering theories, only part of $T(k)$. The assumption that this process was local in wavenumber prompted the name \emph{local energy transfer (LET) theory}. The single-time stationary LET was later extended to the two-time, nonstationary case in a heuristic fashion \cite{McComb78}, and in later years it was developed further as part of the process of reporting various calculations.

Much more recently \cite{McComb17a}, there were two significant steps forward. First, the theory of Edwards gave a clue to the problem. Following the example of the theory of Brownian motion, Edwards clearly envisaged the nonlinear energy transfer in wavenumber as being a Markovian process. This, despite the obvious non-Markovian nature of turbulence which has coupling between the modes. Secondly, the choice of base (or zero-order) operator made by Edwards, was over-determined in order to ensure the Markovian nature of the energy transfer process. When the lowest possible order of operator was chosen instead, the extension of the Edwards SCF theory to the two-time case led directly to the LET theory, as previously derived heuristically, with a response function given by:
\beq
R^{LET}_{\alpha\beta}(\mathbf{k};t,t') =\left\langle\frac{\delta u_\alpha(\mathbf k, t)}{\delta u_{\beta}(\mathbf k, t')}\right\rangle.
\label{letresponse}
\eeq
This relates a change in the velocity field to another change at an earlier time. It may be compared to the DIA form, as given initially by (\ref{inf_resp}), which relates the change in the velocity field to a change in the stirring force.

\subsubsection{Numerical assessment of statistical closure theories}

Even although the theoretical derivations may be intimidating, and not free from controversy, the actual performance of the theories can be understood as readily as any other numerical calculation, such as direct numerical simulation. Accordingly in this section we will give brief introduction to this aspect of the subject.

In 1964 Kraichnan \cite{Kraichnan64b} published numerical calculations of his direct-interaction approximation (DIA) for the free decay of isotropic turbulence from a variety of  initial spectra. At that time he already knew about the problems of his theory at large values of the Reynolds number, but his exploration of its performance at low Reynolds numbers indicated that his theory was a remarkable achievement. It is worth emphasising that his DIA is a theory from first principles, and without adjustable constants, yet gives good qualitative and quantitative results for the free decay of turbulence. At the same time, it should also be said that the experimental position in turbulence at that time did not really permit definitive comparisons. And although the situation in that respect has improved, to some extent this is probably still the case today. Nevertheless, the DIA, even if not entirely correct, was indeed a remarkable achievement.

Kraichnan's own assessment of the DIA, in the  limit of large Reynolds numbers, predicted an inertial-range spectrum of the form $k^{-3/2}$, whereas by the early 1960s the experimental picture had made it clear that the Kolmogorov $k^{-5/3}$ form was correct. His assessment of the failure of DIA under these circumstances prompted a great deal of analysis on his part and ultimately led to, among other things, the \emph{test-field single-time models} and his conclusion that a two-time theory could not be obtained by renormalized perturbation theory in an Eulerian framework. His response to this was to introduce a new hybrid Eulerian-Lagrangian coordinate system, that he called \emph{Lagrangian-history} coordinates. This gave greater freedom of manoeuvre with Lagrangian trajectories, but could be specialised to give an Eulerian form at the end of the calculation. A theory based on this approach (ALHDI) \cite{Kraichnan65} turned out to be not entirely satisfactory; but a subsequent strain-based theory (SBALHDI) performed more satisfactorily \cite{Kraichnan78}.

In 1972, Herring and Kaichnan \cite{Herring72} compared the performance of a number of different theories and models, including the Edwards and Herring self-consistent field theories, the test-field models, and the DIA. They found quite appreciable differences between the various theories, especially in their predictions of the skewness factor. Thus they concluded that even when two theories were close in terms of spectra, the most sensitive indicator of a difference was their prediction of the one-dimensional skewness factor. They did not compute any of these theories at higher Reynolds numbers, but later on computed the Lagrangian-History theories at high Reynolds numbers \cite{Herring79}.
 
In 1984, McComb and Shanmugasundaram \cite{McComb84} published a very full investigation of the LET, as  originally derived in 1978 \cite{McComb78}. They used the same initial spectra and numerical methods as Kraichnan in his assessment of DIA \cite{Kraichnan64b}. The only difference was that they used the $(k,j,\mu=\cos\theta_{kj})$ formulation of Edwards, rather than the $(k,j,l)$ formulation of Kraichnan. This allowed them to make a comparison of the two theories, while using Kraichnan's earlier results as a benchmark. For Taylor-Reynolds numbers up to $R_\lambda \sim 40$ they calculated the free decay of spectra (energy, transfer and dissipation) and found that the LET agreed closely with the DIA. But LET gave higher values of energy transfer and evolved skewness, presumably related to its ability to predict the $-5/3$ spectrum.

The LET was also solved numerically for high Reynolds numbers up to $R_\lambda \sim 533$, and found to agree well with both experimental results and the predictions of the Lagrangian-history theories \cite{Herring79}. In particular, the purely Eulerian LET agreed closely with SBHALDI  and this agreement extended to low values of the Reynolds number as well \cite{McComb84}.

Similar calculations were carried out in 1989 \cite{McComb89} for Taylor-Reynolds numbers in the range $0.5 \leq R_\lambda (t_f)$, where $t_f$ is the final time of the computation. This time the emphasis was on the behaviour of the two-time correlations which were initially plotted for some value of wavenumber against $t_{ref}-t$, where $t_{ref}$ is the final time of the computation. By scaling this, first on the convective time-scale;  and then on the inertial range time-scale, it was established that both scalings could collapse data, with the first being more effective at low wavenumbers, while the second was more effective in the inertial range. It should be borne in mind that these results were for free decay, whereas the results discussed in Section 6 were for stationary turbulence and hence may not be applicable. This leads to an interesting point. In general, results were obtained with $t_{ref}=t_f$, but in one case the results were taken for $t_{ref}<t_f$ and in this one case convective scaling predominateed. Clearly a further investigation to assess the effect of taking results at different times of the decay could be of interest.

Two other general points emerged regarding the comparison between LET and DIA. First, at Reynolds numbers $R_\lambda \leq 5$ the LET gave results that were almost indistinguishable from those of DIA, with the difference between the two tending to zero as $R_\lambda \rightarrow 0$. Secondly, at high Reynolds numbers, DIA did not give the predicted inertial-range $-3/2$ spectrum and instead seemed closer to the $-5/3$ spectrum that the LET gave. This was tested by taking $k^{-3/2}$ as the initial spectrum and it was found that both theories evolved away from this towards $k^{-5/3}$. Evidentally this result does not support Kraichnan's original approximate analysis of the behaviour of DIA at large Reynolds numbers.

For completeness, we should mention the subsequent investigation \cite{McComb92} in which a rederivation of the LET was presented. While still heuristic, this addressed some minor problems with the earlier derivation. This work also applied the theory to passive scalar convection, with quite good results.

\subsubsection{Practical calculations using two-point closures}

In the study of isotropic turbulence, we are less concerned with utility and more with understanding the phenomenon. Nevertheless, most researchers in turbulence have an entirely proper interest in practical applications, so it is worth at least taking a look at this aspect. Unfortunately, in more than sixty years since the pioneering DIA closure, there has been very little interest in applications. A few early instances can be found in \cite{McComb90a} but there are really only two shining examples of applications.

First, there is the work of Orszag, in which he derived the Eddy-damped, Quasi-normal, Markovian closure model (EDQNM) \cite{Orszag70}. This is classed as a model, because it contains an \emph{ad hoc} step which involves an adjustable constant. This is normally chosen to ensure that the model gives a satisfactory value of the Kolmogorov constant. Essentially, Orszag made an analysis of the failure of quasi-normality, and devised ways of putting it right which were based on physical arguments. The end result was a rederivation of the spectral equation due to Edwards \cite{Edwards64}, which we know is compatible with the Kolmogorov spectrum; and an \emph{ad hoc} response equation which was chosen (unlike the Edwards form) to also be compatible with Kolmogorov. There has been some degree of fine tuning of the response equation but this is a successful and (relatively) easily calculated single-time model of turbulence which has been applied to a vast array of non-isotropic turbulence problems from the 1970s to the present day.

Secondly, there is the study of atmospheric and oceanic turbulence, with its relevance to weather and climate change, by Frederiksen and co-workers, using the two-time closures DIA,  Herring's SCF, and  LET (e.g. \cite{Frederiksen04}\nocite{Frederiksen97,Frederiksen04}-\cite{Frederiksen05}). The principal authors have a background in quantum field theory and particle physics so this may have facilitated this line of research, in the sense that they would not have found it as esoteric as most fluid dynamicists certainly do. This activity has extended over three decades with the authors developing their own specialised closures. 

\subsubsection{Lagrangian theories}

In 1981, Kaneda produced a purely Lagrangian theory which was simpler and more elegant than Kraichnan's Lagrangian-history approach. This was termed the Lagrangian Renormalized Approximation or LRA \cite{Kaneda81}. Although simpler that the LHDI theories,  it is still appreciably more complicated than Eulerian theories like the LET. Also, there is the fact that, like Taylor's well known theory of diffusion by continuous movements, translation of the results into the Eulerian coordinate system is a fundamental, unsolved problem. Presumably numerical comparisons rely on the equivalence of Lagrangian and Eulerian spectra for the special case of isotropic turbulence, but if the later results of Tennekes (see Section 6.2 above) are correct then that equivalence does not hold.

Later Kida considered the problem from the point of view of Kraichnan's original perturbation theory \cite{Kida97}, and obtained the same result as Kaneda. Presumably this is another piece of evidence for the equivalence of Kraichnan's original perturbation theory and the conventional form. This was indicated earlier by the fact that Wyld's formalism, when truncated at lowest nontrivial order, yielded the DIA. In passing, we note that Kida's exegesis of the derivation of DIA is much easier to understand than that of Kraichan which is notoriously opaque.

More recently,  Okamura \cite{Okamura18} presented another Lagrangian theory in a paper that also gives a decent review of the the field.

\subsection{Renormalization group mode elimination}

In this section we consider how RG transformations may be implemented in the NSE in wavenumber space. We consider a velocity field $u(k,t)$ on the interval $0 \leq k \leq k_0$, where we employ a simplified notation so that the main ideas may stand out clearly. In the language of theoretical physics, $k_0$ is the ultra-violet cut-off and remains to be chosen. 

In order to introduce the RGT, we consider the NSE on the range $0 < k < k_1$ (explicit scales) with filtered velocity field $u^{-}$; and on $k_1 < k < k_0$ (implicit scales), where $k_1 < k_0$ and the filtered velocity field is $u^{+}$. The terminology refers to large-eddy simulation, and that is just what we are doing, albeit for a rather small range of implicit scales in the dissipation region of the spectrum. Note that the notation for the filtered velocity field is often given as $u^{<}$ and $u^{>}$, respectively, in imitation of  the convention normally used in microscopic physics.

The RGT then involves two steps:
\begin{description}
\item[a] Solve the NSE for $u^{+}$ on the interval $k_1 < k < k_0$. Substitute that solution for the mean effect of the high-$k$ modes into the NSE on $0 < k < k_1$. This results in an increment to the viscosity: $\nu_0 \rightarrow \nu_1 = \nu_0 + \delta\nu_0$.
\item[b] Rescale the basic variables such that the NSE on $0 \leq k \leq k_1$ looks like the original NSE on $0 \leq k \leq k_0$.
\end{description}
This procedure is then repeated for a wavenumber cut-off $k_2 \leq k_1 \leq k_0$, and so on, until a fixed point is reached. The wavenumber bands are chosen by introducing the \emph{spatial rescaling factor} $h$, such that:
\beq
k_1 = hk_0, \quad\mbox{where}\quad 0 < h < 1,
\label{rescale}
\eeq
and this operation is repeated for $k_2 =hk_1$, and so on.

It should be emphasised that the above algorithm is how one implements the RGT \emph{in principle}. There is still the question of how one chooses the initial cut-off wavenumber $k_0$; how one chooses the spatial rescaling factor $h$; and of how one actually carries out the first step of the above algorithm, which strictly involves a nontrivial conditional average. In other words, we have to average combinations of the $u^{+}(k,t)$ while holding the $u^{-}(k,t)$ constant. Denoting such a conditional average by the subscript $C$ on the Dirac brackets, we must have, for example:
\beq
\langle u^-u^+u^+ \rangle_C = u^- \langle u^+u^+ \rangle_C,
\label{condave}
\eeq
where $\langle u^-\rangle_C = u^-$ and $\langle u^+u^+ \rangle_C \neq \langle u^+u^+ \rangle$.

We will discuss some approaches to these questions in the following subsections, where we will make reference to Figure 3.

\subsubsection{FNS theory of stirred fluid motion}

Forster, Nelson and Stephen \cite{Forster77} are generally credited with the first application to the NSE for the particular case of randomly stirred hydrodynamics. They studied the NSE with Gaussian forcing at low wavenumbers and restricted the their analysis to low wavenumbers where cascade effects were excluded by (in effect) choosing $k_0 = \Lambda$ such that they could treat the velocity field as being multivariate normal. As a result, the conditional average of (\ref{condave}) took the trivial form:
\beq
\langle u^-u^+u^+ \rangle_C  = u^- \langle u^+u^+ \rangle_C =  u^- \langle u^+u^+ \rangle,
\label{fns}
\eeq
thus allowing them to take over a theory of magnetism and calculate low-order perturbative corrections due to the nonlinear term. A simplified but detailed treatment can be found in Section 9.3 of \cite{McComb90a}, but the main result is illustrated schematically in Figure 3, where the fixed point is found to be at $k=0$. This Gaussian fixed point is, of course, analogous to the high-temperature fixed point in magnetism; and, like it, is trivial. It should be emphasised that FNS did not claim this to be a theory of turbulence and this point was also made by De Dominicis and Martin \cite{Dedominicis79}. To claim otherwise would be like claiming that the high-temperature fixed point in the Ising model was a satisfactory theory of magnetism, and that there was no need for Wilson's theory in $4-\epsilon$. This would clearly be absurd.

\subsubsection{Iterative averaging}

The first significant application of RG to turbulence was the treatment of the passive scalar problem by Rose \cite{Rose77}, who chose the cutoff wavenumber $k_0$ to be the analogue of the dissipation wavenumber for that problem, and employed the evaluation of the conditional average which was analogous to (\ref{fns}) as an imponderable approximation. The resulting iteratiion reached a fixed point for the renormalized diffusivity which was independent of the initial value but did depend strongly on the choice of spatial rescaling factor.
Two other significant features of this work were that Rose drew attention to the possible use of RG for subgrid modelling and also that the elimination of modes would lead also to an eddy noise term as well as an enhanced viscosity.

The concept of iterative averaging was introduced in 1982 \cite{McComb82}, as an alternative to Reynolds averaging, and which reformulated the statistical equations by averaging over a progressively increasing series of time-scales in real space. Influenced by Rose's theory \cite{Rose77}, the Taylor hypothesis was used to transform the iterative averaging theory to wavenumber space and a calculation carried out which led to a a fixed point corresponding to a  renormalized viscosity. As in Rose's theory, the renormallsed viscosity was found to be independent of the arbitrarily chosen initial viscosity, but did depend on the choice of rescaling factor. Also, the step in (\ref{fns}) was employed as an imponderable approximation.

In time, the choice of the Kolmogorov dissipation wavenumber $k_d = (\varepsilon/\nu_0)^{1/4}$ as the ultraviolet cutoff led to inconsistencies when we tried to calculate the Kolmogorov constant and a new cutoff wavenumber $k_0$ was introduced in 1986 \cite{McComb86}, thus:
\beq
\varepsilon = \int_0^\infty 2\nu_0 k^2 E(k) dk \approx \int_0^{k_0} 2\nu_0 k^2 E(k) dk.
\label{cutoff}
\eeq
The value of $k_0$ depends on how exactly the approximation is evaluated but in practice we found that we had $k_0 = 1.2k_d$. This is a value that is typical of what is taken in practice for the maximum resolved wavenumber in direct numerical simulations. Our first calculation of the Kolmogorov constant was reported in this publication. It varied with the choice of rescaling factor in the range $0.6 \leq h < 1.0$, with a minimum value of around  $\alpha = 1.6$. This of course violates one of the basic tenets of RG, that the value of an observable should be independent of the values taken for the arbitrarily chosen parameters. The introduction of conditional averaging, as we will discuss next, improved this situation.

\subsubsection{Iterative conditional averaging}

The conditional average was introduced in a series of papers \cite{McComb90b}\nocite{McComb92a}-\cite{McComb92b}, and involved three main aspects. These were:
\begin{enumerate}
\item The introduction of a conditional averaging procedure. This involved an extension of classical statistical mechanics through the introduction of a \emph{conditionally sampled ensemble}.
\item The derivation of a hierarchy of governing equations for the conditional averages, with the existing iterative averaging procedures proving suitable for this purpose.
\item Obtaining approximate solutions for conditional averages by introducing the \emph{two-field decomposition} on the right hand side of the governing equations and invoking a hypothesis of \emph{local chaos}.
\end{enumerate}
In this work it was found convenient to introduce the bandwidth $\eta$ of the groups of modes being averaged over in each iteration, and this replaced the rescaling factor $h$, using:
\beq
\eta = 1-h.
\eeq
This procedure allowed us to calculate the Kolmogorov prefactor as $\alpha \sim 1.6$ over a range of arbitrarily chosen band widths.

In later work \cite{McComb01}, a Markovian approximation was replaced by an exact summation, albeit with very little effect of the predictions of the theory. This reference is probably the most detailed account of the current theory, although a later article \cite{McComb06} introduces a minor correction which makes only a very small change to the results. This article also offers a more technical account in  terms of the relationship of the theory to field theoretic approaches and particularly FNS theory \cite{Forster77}. 

To sum up, iterative conditional averaging appears to yield an accurate value for the subgrid energy flux, especially when compared with direct numerical simulation. However, at its present stage of development, it does not take into account the phase coupling associated with a single realisation, nor does it address the eddy noise, as identified by Rose \cite{Rose77}.

\begin{figure} 
\begin{center}
\includegraphics[width=1.0\textwidth, trim=0px 130px 30px 0px,clip]{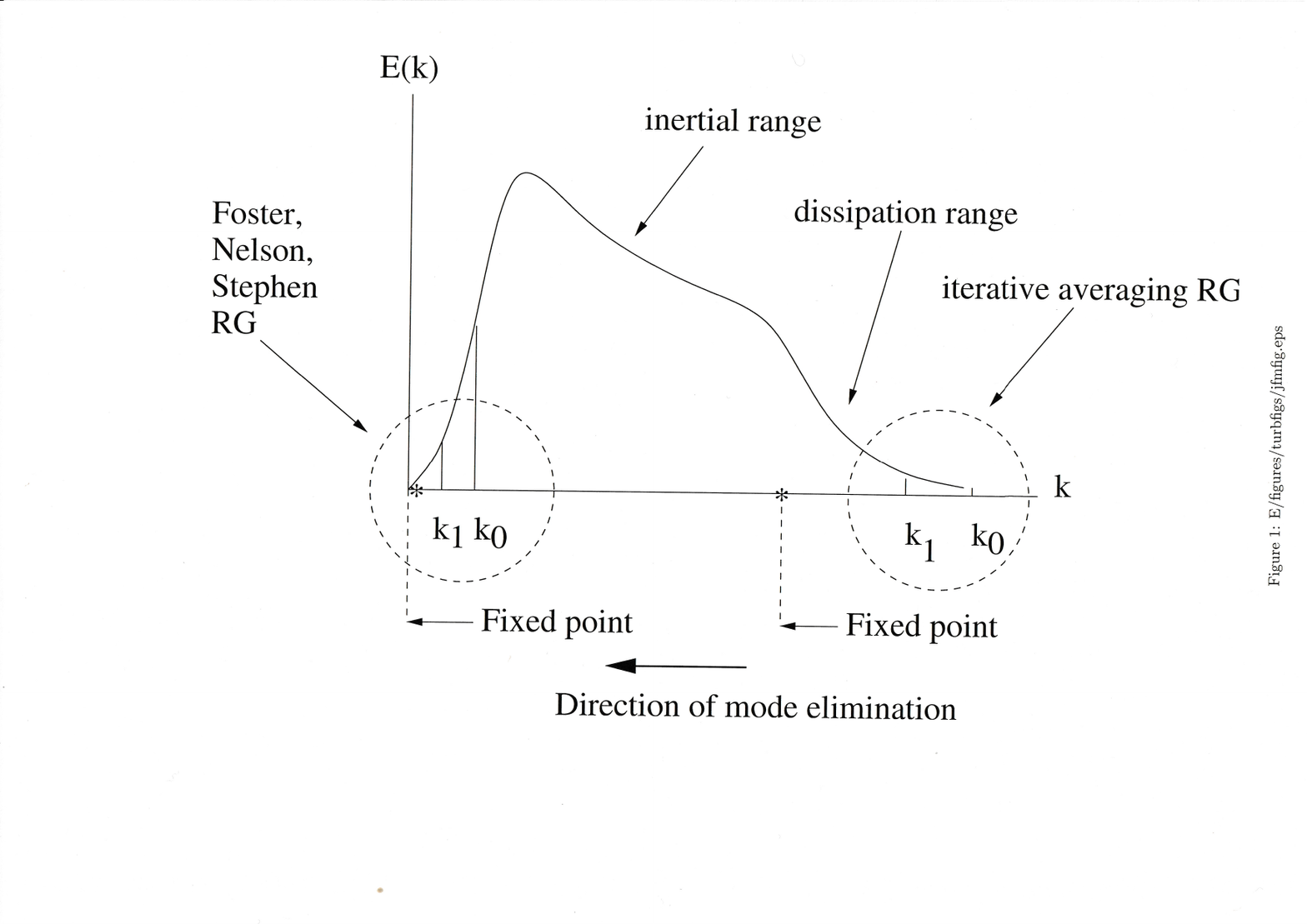}
\end{center} 
\caption{\small  Schematic view of the energy spectrum showing the application of renormalization group to the Navier-Stokes equation for stirred fluid motion and for turbulence.} 
\label{fig3} 
\end{figure}

\subsection{Implications of the failure of the Direct Interaction Approximation}

The title of this subsection is a misnomer, but it is also realistic. First, it is a misnomer because it is the failure of all three cognate pioneering theories which have significant implications; and in particular the failure of the SCF theory of Edwards that holds lessons for us now. Secondly, it is realistic because a significant fraction of the turbulence community believe that the failure of Kraichnan's DIA means that renormalized perturbation theory cannot work in Eulerian coordinates and hence one must try Lagrangian theories. The persistence of this belief despite the success of the LET theory in curing the problems with the single-time Edwards theory in 1974 \cite{McComb74} and with Kraichnan's two-time theory in 1978 \cite{McComb78} seems rather puzzling. What we can conclude is that Eulerian perturbation theory that begins with Kraichnan's starting point of a relationship between the stirring forces and the velocity field probably cannot succeed.

In the 1960s, when the present author began working as a graduate student, there was a great deal of interest in this disappointing failure. Generally it seemed to be something to do with a problem at low wavenumbers, with Edwards showing that a consideration of  his theory in the limit of infinite Reynolds numbers led to an infra-red divergence in the response equation. The much more complicated analysis of Kraichnan's, much more complicated, theory, suggested that the convective time-scale, which one would associate with low-$k$ modes, had a dominant effect.

It took until 2017 \cite{McComb17a} for it to be realized that Kraichnan's DIA and the Edwards SCF had something more in common. That is, they each represented a Markovian approximation to the turbulence energy cascade in wavenumber. In fact, in the Edwards SCF, this was manifest in that he drew an analogy with the random walk of Brownian motion. This conclusion applies to Herring's SCF by extension. It should also be noted that this Markovian approximation in wavenumber should not be confused with the Markovian approximation in the EDQNM model which involves a Markovian approximation in time.

The implication of all this, is that starting from equation (\ref{inf_resp}) leads both to a Markovian representation of the energy cascade and to a failure to yield the Kolmogorov spectrum. As both the Wyld and MSR formalisms have that as their starting point, and as both at lowest nontrivial order are the same as the DIA, it follows that these formalisms are not representative of turbulence, any more than the three pioneering theories are. In particular, any formalism or theory which begins by assuming the MSR action is flawed from the outset and is subject to imponderable errors e.g. \cite{Tomassini97}.

\section{Concluding remarks}

It has been known since at least the 1930s that turbulent shear flows exhibit asymptotic behaviour with increasing Reynolds number. To take pipe flow as an example, plotting the mean velocity profile $\bar{U}_1/U_c$, where the mean flow is in the $x_1$ direction and $U_c$ is the velocity at the centreline, against the distance from the wall divided by the radius of the pipe, yields a family of curves which tend to a limiting form as the Reynolds number is increased. In the case of isotropic turbulence, it has long been suspected that the dimensionless dissipation rate also tends to a limiting form, and the recent theory leading to equation (\ref{ourdiss}) has established that this is the case. As we have pointed out, this relationship is either exact or asymptotically exact, but in either case it leads to an asymptotic value of the dimensionless dissipation which we denote by $C_{\varepsilon,\infty}$.

We should note that this result depends on the existence of the dimensionless structure functions in the limit of large Reynolds numbers. This would seem to  be a reasonable assumption. Also, the existence of this asymptotic behaviour corresponds to the Kolmogorov (1941) picture, in which the inertial transfer becomes equal to the dissipation. As we have pointed out earlier, this basic picture is more easily understood in wavenumber ($k$) space, than in real ($x$) space.

Turning more specifically to the second question in the title of this article, we would argue that there are two principal motivations for studying isotropic turbulence. First, there is the desire to understand the phenomenon of turbulence in its own right, as a branch of physics. Indeed Taylor proposed the study of isotropic turbulence (especially in the Fourier wavenumber representation) as a branch of statistical mechanics. Nowadays we would see that as a branch of many-body physics or statistical field theory. Secondly, for many it is seen as a starting point for the consideration of more general, practical problems. This means that the underlying emphasis is on moving on to applications.

There is no inherent reason why these two approaches should be mutually exclusive. Nevertheless, it is difficult for anyone working in turbulence to escape the feeling that the subject is of great practical importance, and this can be inhibiting when one is dealing with fundamentals. Just as pure mathematicians need to study mathematics for its own sake, so also do theoretical physicists have to study mathematical problems in physics for their own sake, albeit with an input of \emph{physical} understanding in order to solve what is mathematically insoluble. At the same time, the publication of scientific papers and the award of research grants can be at the mercy of those with a utilitarian attitude to the subject, who apparently do not wish to see what they believe is `the existing fundamentals' changed in any way. We should bear in mind, the attitude of the great theoretical physicist Richard Feynmann, who said:
\begin{quote}
`I would rather have questions that can't be answered than answers that can't be questioned.'
\end{quote}
It is worth bearing in mind that Feynmann unlocked the creativity which would lead to his Nobel Prize by deciding only to work on what was fun!


\end{document}